\documentclass[man,floatsintext, donotrepeattitle, 11pt]{apa7}
\usepackage{lipsum}
\usepackage{amsmath}
\usepackage[utf8]{inputenc}
\usepackage{amssymb}
 \usepackage{hyperref}
\allowdisplaybreaks
\usepackage{float}
\usepackage[font=small,labelfont=bf,margin=1in]{caption}
\usepackage{tikz}
\usetikzlibrary{decorations.markings}
\usetikzlibrary{arrows.meta, positioning}
\usetikzlibrary{decorations.pathmorphing}
\usetikzlibrary{shapes.geometric, calc,decorations.pathreplacing}
\usetikzlibrary{backgrounds}
\usetikzlibrary{fit}
\usetikzlibrary{arrows}
\tikzset{
	arrow/.pic={\path[tips,every arrow/.try,->,>=#1] (0,0) -- +(.1pt,0);},
	pics/arrow/.default={triangle 90}
}

\usepackage[american]{babel}

\usepackage{csquotes}
\usepackage[style=apa,natbib=true,sortcites=true,sorting=nyt,backend=biber, doi = false]{biblatex}
\DeclareLanguageMapping{american}{american-apa}
\usepackage[stable]{footmisc}
\usepackage[font=small,labelfont=bf,margin=1in]{caption}

\newcommand{\bP}{\mathbb{P}} 
\newcommand{\bE}{\mathbb{E}}

\newcommand{\bs}{\boldsymbol}

\usepackage{listings}
\usepackage{xcolor}

\definecolor{codegreen}{rgb}{0,0.6,0}
\definecolor{codegray}{rgb}{0.5,0.5,0.5}
\definecolor{codepurple}{rgb}{0.58,0,0.82}
\definecolor{backcolour}{rgb}{0.95,0.95,0.92}

\lstdefinestyle{Rstyle}{
    backgroundcolor=\color{white},   
    commentstyle=\color{codegreen},
    keywordstyle=\color{black},
    numberstyle=\tiny\color{codegray},
    stringstyle=\color{codepurple},
    basicstyle=\ttfamily\footnotesize,
    breakatwhitespace=false,         
    breaklines=true,                 
    captionpos=b,                    
    keepspaces=true,                 
    numbers=left,                    
    numbersep=5pt,                  
    showspaces=false,                
    showstringspaces=false,
    showtabs=false,                  
    tabsize=2,
    language=R,
  escapeinside={(*@}{@*)},
  moredelim=**[is][\color{red}]{@@}{@@},
  moredelim=**[is][\color{gray}]{<<}{>>}
}

\title{From Structural Equation Modeling to Targeted Learning: A Tutorial Introduction to Targeted Maximum Likelihood Estimation for SEM Researchers}
\shorttitle{From Path Coefficients to Targeted Estimands}

\authorsnames[1,2,3,4,3,5]{
  Junjie Ma,
  Xiaoya Zhang,
  Guangye He,
  Yuting Han,
  Ting Ge,
  Feng Ji,
}

\authorsaffiliations{
  {Department of Statistical Sciences, University of Toronto, Toronto, Canada},
  {Department of Family, Youth and Community Sciences, University of Florida, Gainesville, U.S.A.},
  {School of Social and Behavioral Sciences, Nanjing University, Nanjing, China},
  {School of Psychological and Cognitive Sciences, Beijing Language and Culture University, Beijing, China},
  {Department of Applied Psychology and Human Development, University of Toronto, Toronto, Ontario, Canada}
}

\leftheader{*}

\abstract{Structural equation modeling (SEM) and path analysis have long been central tools for studying complex causal relationships in the social and behavioral sciences, yet their reliance on parametric assumptions can lead to biased inference under model misspecification. To bridge traditional SEM with modern causal machine learning, this paper introduces targeted maximum likelihood estimation (TMLE), a doubly robust framework built on nonparametric structural equation modeling. We formally connect TMLE to classical path analysis, showing that standard SEM estimators arise as special cases of TMLE under restrictive parametric specifications and that both approaches can estimate common causal quantities such as direct, indirect, and total effects. Through simulation studies under both correctly specified and misspecified models, we demonstrate that while the two methods perform similarly when models are correctly specified, TMLE consistently achieves lower bias, reduced mean squared error, and improved confidence interval coverage when parametric assumptions are violated. We further illustrate these differences using an applied mediation analysis examining the role of poverty in access to high school education, where path analysis suggests a significant direct effect, whereas TMLE does not, highlighting the practical consequences of robustness in causal inference. Overall, this tutorial offers SEM researchers a conceptual and practical introduction to targeted learning, providing guidance on leveraging TMLE to enhance causal analysis beyond traditional parametric frameworks.}


\begin{document}
\maketitle

\section{Introduction}

Structural equation modeling (SEM; \citealp{bollen-2014}) has long served as a central analytical framework in applied statistics, the social sciences, psychology, and related disciplines (e.g., \citealp{long-2023, szaflarski-2019, hermstad-2010}), enabling researchers to represent complex relationships among observed and latent variables. As a special case of SEM involving only observed variables, path analysis provides an accessible approach for examining mediational mechanisms through the estimation of direct and indirect effects \citep{curran-2003, gunzler-2013}. Beyond its conceptual appeal, SEM benefits from extensive software support, including implementations in \texttt{lavaan} \citep{rosseel-2012} for \texttt{R}, as well as widely used platforms such as \texttt{Stata} and \texttt{Mplus}, which have facilitated its broad applications in empirical research.

A wide range of research questions in the social and behavioral sciences can be addressed through the estimation of path coefficients within the SEM framework \citep[e.g.,][]{wood-2023,natukunda-2024}. Under specified structural equations and distributional assumptions on the disturbance terms, model parameters are typically estimated via maximum likelihood. Statistical inference is subsequently conducted using likelihood-based tests or resampling approaches such as the bootstrap \citep{bollen-2014}. However, since the true data-generating mechanism is typically unknown in practice, the specified likelihood may fail to accurately represent the underlying process \citep{curran-2003}. When model misspecification occurs, the desirable properties of SEM estimators, including consistency, efficiency, and valid uncertainty quantification, can be compromised, leading to biased estimates and unreliable inference \citep{kaplan-1988, yuan-2003}.

Path diagrams and the corresponding path coefficients are often interpreted causally, and under certain conditions they can correspond to causal effects defined within the Neyman–Rubin potential outcomes framework \citep[e.g.,][]{bollen-2013, pearl-2009}. This connection provides an appealing link between causal inference and traditional parameter estimation in SEM. Nevertheless, such equivalence depends on correct model specification, including linearity and appropriate distributional assumptions. To relax these restrictions and reduce sensitivity to parametric assumptions, nonparametric structural equation models (NPSEM; \citealp{pearl-2009-1}) represent relationships among variables using flexible, nonparametric functions. Within this framework, path diagrams are reformulated as directed acyclic graphs (DAGs), and causal effects are identified as functionals of the observed data distribution rather than as specific path coefficients.

Building on the NPSEM framework, targeted maximum likelihood estimation (TMLE), introduced by \citet{van-der-laan-2011}, provides a modern semiparametric approach to causal inference that integrates machine learning with statistical estimation. TMLE proceeds in two stages: an initial data-adaptive estimation of relevant components of the data-generating distribution, followed by a targeted update that solves the estimating equation associated with the efficient influence function (EIF) \citep{gruber-2009}. In practice, TMLE is often paired with the Super Learner algorithm \citep{van-der-laan-2006}, an ensemble method that optimally combines multiple prediction algorithms to improve estimation accuracy. Under regularity conditions, TMLE achieves asymptotic efficiency and enables valid uncertainty quantification via the influence function, offering both theoretical guarantees and computational advantages over resampling-based approaches \citep{gruber-2009, van-der-laan-2011}.

TMLE relaxes the model assumptions, thus providing a robust alternative to traditional parametric SEM and path analysis, particularly in settings where model assumptions are likely to be violated. By leveraging flexible machine learning estimators while preserving causal interpretability, TMLE mitigates the risks associated with misspecification and extends the scope of causal analysis beyond parametric frameworks. Although several TMLE tutorials for ATE exist \citep[e.g.,][]{luquefernandez-2018,frank-2023}, there is still no accessible tutorial for mediation analysis and for behavioral and social science researchers that frames TMLE through the lens of parametric SEM, which leaves many SEM users without a clear path to adopt methods that are more robust to misspecification compared to the parametric path analysis. In this paper, we provide a tutorial introduction to TMLE for researchers familiar with SEM and path analysis, emphasizing shared conceptual foundations and showing how parametric assumptions in classical SEM estimators can be relaxed to gain robustness. Through both theoretical connections and empirical demonstrations, we aim to show that TMLE offers improved robustness and performance in a wide range of causal inference problems.

The remainder of the paper is organized as follows. In Section~2, we review the core frameworks of SEM and TMLE. Section~3 introduces fundamental causal effects and the assumptions required for their identification, followed by illustrations of how path analysis and TMLE can be applied to estimate these effects, including mediation analysis. In Section~4, we present simulation studies under correctly specified models as well as various forms of misspecification, including omitted interactions, nonlinear relationships, and non-normal disturbances, to compare the performance of SEM and TMLE. Section~5 provides an applied example analyzing the mediating role of poverty in access to high school education using both approaches. Section~6 concludes with a discussion of practical implications and future directions.
\section{Review of SEM and TMLE}\label{Intro}

\subsection{Structural Equation Modeling and Path Analysis}
\setlength{\abovedisplayskip}{4pt}
\setlength{\belowdisplayskip}{4pt}

Structural equation modeling (SEM) provides a flexible parametric framework for representing complex relationships among observed and latent variables. The presentation here closely follows \citet{curran-2003}, with minor modifications for notational consistency and emphasis on SEM as a parametric modeling approach that is frequently interpreted causally in applied research. In its general form, SEM consists of a structural submodel describing relations among latent variables and a measurement submodel linking latent constructs to observed indicators.

The structural submodel is typically specified as
\begin{align*}
\boldsymbol{\eta} = \boldsymbol{\mu} + \boldsymbol{\beta}\boldsymbol{\eta} + \boldsymbol{\zeta},
\end{align*}
where $\boldsymbol{\eta} \in \mathbb{R}^k$ denotes the vector of latent variables, $\boldsymbol{\mu} \in \mathbb{R}^k$ represents latent intercepts, $\boldsymbol{\beta}$ is a $k \times k$ matrix of regression coefficients encoding directional relations among latent factors, and $\boldsymbol{\zeta}$ is a disturbance vector assumed to be multivariate normal with mean $\boldsymbol{0}$ and covariance matrix $\boldsymbol{\Psi}$.

The measurement submodel links latent variables to observed outcomes,
\begin{align*}
\boldsymbol{y} = \boldsymbol{\nu} + \boldsymbol{\Lambda}\boldsymbol{\eta} + \boldsymbol{\epsilon},
\end{align*}
where $\boldsymbol{y} \in \mathbb{R}^p$ is the vector of observed variables, $\boldsymbol{\nu}$ denotes measurement intercepts, $\boldsymbol{\Lambda}$ is the $p \times k$ factor loading matrix, and $\boldsymbol{\epsilon}$ denotes measurement errors with mean $\boldsymbol{0}$ and covariance matrix $\boldsymbol{\Sigma}_{\epsilon}$. We further assume $\mathbb{E}[\boldsymbol{\zeta}]=\mathbb{E}[\boldsymbol{\epsilon}]=\boldsymbol{0}$ and $\text{Cov}(\boldsymbol{\zeta},\boldsymbol{\epsilon})=\boldsymbol{0}$.

Substituting the structural equation into the measurement model yields
\begin{align*}
\boldsymbol{y} = \boldsymbol{\nu} + \boldsymbol{\Lambda}\boldsymbol{B}\boldsymbol{\mu} + \boldsymbol{\Lambda}\boldsymbol{B}\boldsymbol{\zeta} + \boldsymbol{\epsilon},
\end{align*}
where $\boldsymbol{B} = (\boldsymbol{I}_k - \boldsymbol{\beta})^{-1}$. Consequently, the implied mean and covariance of $\boldsymbol{y}$ are
\begin{align*}
\mathbb{E}[\boldsymbol{y}] &= \boldsymbol{\nu} + \boldsymbol{\Lambda}\boldsymbol{B}\boldsymbol{\mu} := \boldsymbol{\mu}_y, \\
\text{Cov}(\boldsymbol{y}) &= \boldsymbol{\Lambda}\boldsymbol{B}\boldsymbol{\Psi}\boldsymbol{B}^{\top}\boldsymbol{\Lambda}^{\top} + \boldsymbol{\Sigma}_{\epsilon} := \boldsymbol{\Sigma}_y.
\end{align*}

Under multivariate normality, parameters are typically estimated by maximizing the induced likelihood function. For a sample of size $n$ with sample mean $\bar{\boldsymbol{y}}$ and sample covariance matrix $\boldsymbol{S}$, the log-likelihood is
\begin{align*}
\ell(\boldsymbol{\theta}) = -\frac{n}{2}\Big[ \log|\boldsymbol{\Sigma}_y| + \text{tr}(\boldsymbol{\Sigma}_y^{-1}\boldsymbol{S}) + (\bar{\boldsymbol{y}}-\boldsymbol{\mu}_y)^{\top}\boldsymbol{\Sigma}_y^{-1}(\bar{\boldsymbol{y}}-\boldsymbol{\mu}_y) + p\log(2\pi) \Big],
\end{align*}
where $\boldsymbol{\theta}$ collects all free parameters. Maximizing this likelihood (equivalently minimizing its negative) yields $\hat{\boldsymbol{\theta}}_{\text{ML}}$. Under regularity conditions, including the correct specification of the structural equations, $\sqrt{n}(\hat{\boldsymbol{\theta}}_{\text{ML}}-\boldsymbol{\theta})$ is asymptotically normal with covariance given by the inverse Fisher information matrix, enabling Wald-type inference and likelihood ratio testing. When estimands are smooth transformations of path coefficients (e.g., indirect effects), uncertainty is usually quantified via the Delta method or resampling-based procedures \citep{bollen-2013}.
If the model is misspecified, the maximum likelihood estimator can lose key properties (e.g., consistency and correctly calibrated standard errors), leading to biased or otherwise unreliable estimation and inference.

When all variables are observed with no latent variables involved, SEM reduces to a system of simultaneous linear regression equations commonly referred to as \emph{path analysis}. In this setting, the implied mean and covariance are derived solely from the specified structural equations, and parameters can still be estimated by maximum likelihood \citep{de-stavola-2014}. Path analysis then serves as a canonical fully parametric SEM framework in applied research, particularly for mediation and pathway analysis. For clarity and consistency with the tutorial goals of this paper, we will henceforth use the term SEM primarily to refer to models involving only observed variables unless otherwise noted.

\subsection{Targeted Learning: From NPSEM to TMLE}

\subsubsection{An Overview of Targeted Learning}
In contrast to parametric SEM and path analysis, which rely on restrictive functional and distributional assumptions, targeted learning provides a semiparametric framework for estimating causal effects and other statistical parameters under minimal modeling constraints. Targeted maximum likelihood estimation (TMLE) is built upon the nonparametric structural equation modeling (NPSEM) framework, in which causal relationships are represented via directed acyclic graphs and causal parameters are identified as functionals of the observed-data distribution rather than as parametric path coefficients. Once a causal model and an estimand are specified, TMLE combines flexible machine learning estimation with a principled targeting update to yield robust and efficient inference.

Conceptually, as shown in Figure~\ref{fig:tmle_flow}, TMLE proceeds in four steps: (1) specify the statistical model and target estimand; (2) obtain initial data-adaptive estimates of the relevant components of the data-generating distribution; (3) ``target'' these estimates to solve the efficient influence function (EIF) estimating equation for the estimand; and (4) conduct inference using influence-function-based standard errors. The targeting step is what distinguishes TMLE from generic plug-in machine learning estimators, ensuring local efficiency and asymptotic linearity under mild regularity conditions.

\begin{figure}[h!]
\centering
\begin{tikzpicture}[
    node distance=1.2cm,
    every node/.style={font=\small},
    block/.style={
        rectangle,
        draw,
        rounded corners,
        align=center,
        minimum width=8.2cm,
        minimum height=1.15cm
    },
    line/.style={-latex, thick}
]
\node[block] (step1) {{\bf Step 1:} Define model \& target parameter\\
\footnotesize Specify the data structure and causal/statistical model;\\
define the estimand (e.g., causal effect).};

\node[block, below=of step1] (step2) {{\bf Step 2:} Initial estimation (often Super Learner)\\
\footnotesize Use cross-validated learning to estimate the\\
relevant parts of the distribution (e.g., $Q$, $g$).};

\node[block, below=of step2] (step3) {{\bf Step 3:} Targeting step (TMLE)\\
\footnotesize Construct clever covariates and update the initial\\
estimate via a fluctuation model to target the estimand.};

\node[block, below=of step3] (step4) {{\bf Step 4:} Inference \& interpretation\\
\footnotesize Use influence-function-based (or resampling-based) standard errors\\
for confidence intervals and interpret the estimated parameter.};

\draw[line] (step1) -- (step2);
\draw[line] (step2) -- (step3);
\draw[line] (step3) -- (step4);
\end{tikzpicture}
\caption{Flowchart of the TMLE procedure, adapted from \citet{van-der-laan-2011}.}
\label{fig:tmle_flow}
\end{figure}

Formally, following \citet{gruber-2010}, let $\mathcal{M}$ denote a semiparametric statistical model containing the true but unknown distribution $F_0 \in \mathcal{M}$. Let the parameter of interest be $\Psi(F_0)$, where $\Psi: \mathcal{M} \rightarrow \mathbb{R}$, and let $O_1,\ldots,O_n \stackrel{i.i.d.}{\sim} F_0$. In many applications, $\Psi(F_0)$ depends only on a component $Q_0 = Q(F_0)$ of $F_0$, while additional components $g_0$ act as nuisance parameters under an orthogonal factorization. We assume $\Psi(F_0)=\Psi(Q_0)$.

The first step in TMLE is to obtain an initial estimator $Q_n^0$ of $Q_0$ by minimizing an empirical loss:
\begin{align*}
Q_0 &= \arg\min_{Q \in \mathcal{Q}} \mathbb{E}_{O \sim F_0}\big[\mathcal{L}_Q(O)\big],
\quad \mathcal{Q}:=\{Q(F):F \in \mathcal{M}\},\\
Q_n^0 &= \arg\min_{Q \in \mathcal{Q}} \frac{1}{n}\sum_{i=1}^n \mathcal{L}_Q(O_i),
\end{align*}
where common choices of $\mathcal{L}_Q$ include squared-error loss and negative log-likelihood depending on the outcome type.

Next, obtain an estimator $g_n$ of the nuisance parameter $g_0$. Given $(Q_n^0,g_n)$, TMLE defines a parametric fluctuation submodel $Q_{n,g_n}(\epsilon)$ satisfying
\begin{align*}
\left.\frac{d}{d\epsilon}\mathcal{L}\big(Q_{n,g_n}(\epsilon)\big)(O)\right|_{\epsilon=0}
= D^*(Q_n^0,g_n)(O),
\end{align*}
where $D^*$ is the efficient influence function of $\Psi$.

The fluctuation parameter is estimated by empirical risk minimization,
\begin{align*}
\epsilon_n = \arg\min_{\epsilon} \frac{1}{n}\sum_{i=1}^n \mathcal{L}\big(Q_{n,g_n}(\epsilon)\big)(O_i),
\end{align*}
yielding the targeted estimate $Q_n^* = Q_{n,g_n}(\epsilon_n)$. This update may be iterated until convergence. The final targeted estimate satisfies the empirical EIF estimating equation,
\begin{align*}
\frac{1}{n}\sum_{i=1}^n D^*(Q_n^*,g_n)(O_i)=0,
\end{align*}
which implies the asymptotic distribution
\begin{align*}
\sqrt{n}\big(\Psi(Q_n^*)-\Psi(F_0)\big)\xrightarrow{d}\mathcal{N}\big(0,\text{Var}(D^*)\big),
\end{align*}
allowing construction of valid confidence intervals using influence-function-based standard errors. {  In contrast to SEM and path analysis, TMLE does not require specifying a fully parametric model for the structural relations. Instead, it targets the estimands directly as functionals of the observed-data distribution within a semiparametric and NPSEM framework, yielding greater robustness to model misspecification.}

TMLE has been extended to a wide range of estimands that admit asymptotically linear representations \citep[e.g.,][]{gruber-2010, stitelman-2012, hejazi-2022}. For simplicity, we use TMLE and \emph{targeted learning} interchangeably to refer to this general framework.

\subsubsection{The Super Learner for Data-Adaptive Estimation}
As mentioned, the performance of TMLE depends critically on the initial estimates $Q_n^0$ and $g_n$. Although parametric regressions can be used, targeted learning typically employs flexible machine learning methods to reduce sensitivity to misspecification. Since no single learning algorithm is uniformly optimal, TMLE is often paired with the Super Learner algorithm \citep{van-der-laan-2007}, an ensemble approach with strong theoretical guarantees.

Consider data $\{(X_i,Y_i)\}_{i=1}^n$ drawn from an unknown distribution $F_0 \in \mathcal{M}$. The goal is to construct a predictor $f(X)$ that minimizes the expected risk
\begin{align*}
R(f)=\mathbb{E}_{F_0}\big[\mathcal{L}(Y,f(X))\big],
\end{align*}
where $\mathcal{L}$ is a loss function (e.g., squared error for regression or log-loss for classification). Super Learner forms a weighted ensemble of $M$ user-specified candidate learners $\{f_m\}_{m=1}^M$,
\begin{align*}
f_{\text{SL}}(X) = \sum_{m=1}^M \hat{w}_m f_m(X),
\end{align*}
where weights satisfy $\hat{w}_m \ge 0$ and $\sum_{m=1}^M \hat{w}_m = 1$.

\begin{figure}[h]
\centering
\includegraphics[width=0.9\linewidth]{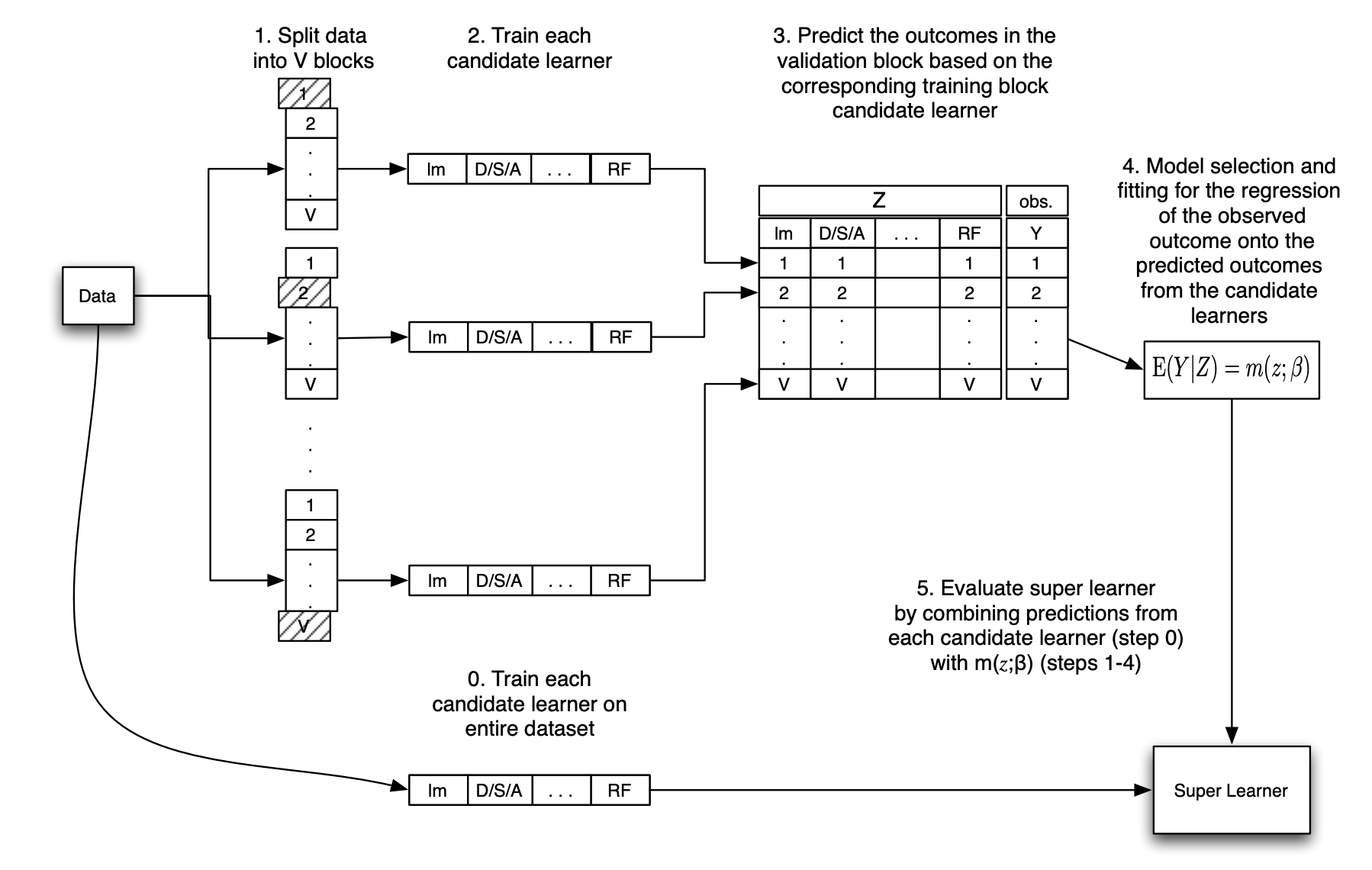}
\caption{Illustration of the Super Learner algorithm. From \citet{van-der-laan-2007}.}
\label{fig:superlearner}
\end{figure}

Weights are estimated using $V$-fold cross-validation. Partition the data into folds $\{D_v\}_{v=1}^V$. For each fold $v$, train learner $m$ on $D_{-v}=D\setminus D_v$ and predict on $D_v$, producing $Z_{i,m}=f_m^{-v}(X_i)$ for $i\in D_v$. These predictions form a level-one matrix $Z\in\mathbb{R}^{n\times M}$. A meta-learner, commonly nonnegative least squares, solves
\begin{align*}
\hat{w}=\arg\min_{w\in\mathcal{W}}\frac{1}{n}\sum_{i=1}^n
\mathcal{L}\Big(Y_i,\sum_{m=1}^M w_m Z_{i,m}\Big),
\end{align*}
where $\mathcal{W}=\{w: w_m\ge0, \sum_{m=1}^M w_m=1\}$. An intuitive illustration is given in Figure~\ref{fig:superlearner}. Under mild conditions, Super Learner achieves an Oracle inequality, asymptotically performing as well as the best combination of learners in the library \citep{van-der-laan-2007}.

Modern developments and implementations of Super Learner support a broad set of modern machine learning and deep learning algorithms \citep[e.g.,][]{li2025targeteddeeparchitecturestmlebased,young-2018}, which improves flexibility across applications. In practice, the choice of base learners and meta-learner influences performance and should be guided by the outcome type, sample size, and expected complexity of the underlying relationships \citep{phillips-2023}.

\section{SEM and TMLE Approaches to Causal Inference} \label{main}

Parametric SEM and path analysis are widely used to quantify associations and mediational pathways, and under additional identifying assumptions, they can be interpreted causally. Common causal estimands studied in SEM-based applications include average treatment effects (ATE), conditional average treatment effects (CATE), and mediation effects such as natural direct and indirect effects \citep{gunzler-2013, bollen-2013}. In this section, we review these estimands under the potential outcomes framework and show how both the parametric path-analysis approach and the Targeted Learning approach can be used for estimation. The key message being consistently emphasized is that, under correct parametric specification, path-analysis estimators and TMLE can be asymptotically equivalent, whereas TMLE offers improved robustness when parametric working models are potentially misspecified.

\subsection{Average Treatment Effects}

\subsubsection{Estimands and identification assumptions}
Suppose we observe i.i.d.\ data $O_i=(Y_i,A_i,\bs X_i)$ for $i=1,\ldots,n$, where $Y$ is an outcome, $A\in\{0,1\}$ is a binary treatment or exposure, and $\bs X\in\mathcal{X}\subseteq\mathbb{R}^p$ denotes pre-treatment covariates. Let $Y(1)$ and $Y(0)$ denote potential outcomes under $A=1$ and $A=0$, respectively. The average treatment effect (ATE) is defined as
\begin{align*}
\psi_{\textup{ATE}} = \mathbb{E}\!\left[Y(1)-Y(0)\right].
\end{align*}
To capture heterogeneity, the conditional average treatment effect (CATE) for a subgroup $B\subset\mathcal{X}$ is defined as
\begin{align*}
\psi_{\textup{CATE}}(B) = \mathbb{E}\!\left[Y(1)-Y(0)\mid \bs X\in B\right].
\end{align*}

Identification from observed data typically relies on the following assumptions \citep{rosenbaum-1983}: 
(i) \emph{consistency/SUTVA}: $Y = A\,Y(1)+(1-A)\,Y(0)$; 
(ii) \emph{no unmeasured confounding} (conditional exchangeability): $(Y(1),Y(0)) \perp A \mid \bs X$; and 
(iii) \emph{positivity}: for some $\epsilon>0$, $\mathbb{P}(A=a\mid \bs X=\bs x)>\epsilon$ for $a\in\{0,1\}$ and almost all $\bs x\in\mathcal{X}$. 
Under these assumptions, the ATE can be written as
\begin{align*}
\psi_{\textup{ATE}}
= \mathbb{E}_{\bs X}\!\left[\mathbb{E}(Y\mid A=1,\bs X)-\mathbb{E}(Y\mid A=0,\bs X)\right].
\end{align*}

\subsubsection{Path analysis as a parametric estimator of ATE and CATE}
\setlength{\abovedisplayskip}{4pt}
\setlength{\belowdisplayskip}{4pt}

A common SEM and path-analysis approach is to impose a parametric structural model in which the treatment effect is represented by a path coefficient. For concreteness, consider the linear structural causal model \citep{pearl-2009}
\begin{align*}
&(U_{\bs X},U_A,U_Y)\sim F_U,\\
&\bs X = U_{\bs X},\\
&A = g_A(\bs X,U_A),\\
&Y = \gamma A + \bs\beta^\top \bs X + U_Y,
\end{align*}
where $(U_{\bs X},U_A,U_Y)$ are mutually independent exogenous variables and $g_A$ is a known assignment mechanism (e.g., logistic). Under conditional exchangeability, the path coefficient $\gamma$ equals the ATE:
\begin{align*}
\psi_{\textup{ATE}}
&= \mathbb{E}_{\bs X}\!\left[\mathbb{E}(Y\mid A=1,\bs X)-\mathbb{E}(Y\mid A=0,\bs X)\right]\\
&= \mathbb{E}_{\bs X}\!\left[(\gamma+\bs\beta^\top \bs X)-(\bs\beta^\top \bs X)\right]
= \gamma.
\end{align*}
Hence, estimating the ATE reduces to estimating $\gamma$. Under correct model specification, maximum likelihood (or equivalently least squares for linear Gaussian models) yields an unbiased, consistent, and asymptotically normal estimator, and inference about $H_0:\gamma=0$ is commonly performed using Wald or likelihood ratio tests.

To allow heterogeneous effects, one can extend the structural model to include treatment--covariate interactions. For example, if the treatment interacts with $X_k$,
\begin{align*}
Y = \gamma A + \bs\beta^\top \bs X + \tau\,A X_k + U_Y,
\end{align*}
then the CATE for subgroup $B\subset\mathcal{X}$ becomes
\begin{align*}
\psi_{\textup{CATE}}(B)
&= \mathbb{E}\!\left[Y(1)-Y(0)\mid \bs X\in B\right]\\
&= \gamma + \tau\,\mathbb{E}\!\left[X_k\mid \bs X\in B\right].
\end{align*}
If $X_k$ is constant on $B$ (e.g., $X_k=x_k$), then $\psi_{\textup{CATE}}(B)=\gamma+\tau x_k$. Thus, estimating the CATE is then equivalent to estimating a linear transformation of the path coefficients. Estimation and inference follow from standard SEM theory under correct specification, with uncertainty for CATE obtained via the Delta method or standard linear combinations of asymptotically normal estimators.\\
Identifying causal effects as functionals of the joint distribution relies primarily on the causal assumptions encoded in the causal diagram. However, expressing those causal effects in terms of path coefficients typically requires additional parametric restrictions on the data-generating process (e.g., linearity), which are often misspecified. Under misspecification, the estimation may be biased and converge to the pseudo-true parameters. A natural alternative is then to formulate the data-generating process as a NPSEM and to target the identified functionals directly, an approach that leads naturally to TMLE.

\subsubsection{TMLE for ATE and CATE under an NPSEM}
As mentioned earlier, interpreting and deriving the ATE and CATE as path coefficients hinges on correctly specified parametric models. To relax the strong parametric assumptions imposed by SEM, targeted learning instead begins with a nonparametric structural equation model (NPSEM) \citep{pearl-2009}:
\begin{align*}
&(U_{\bs X},U_A,U_Y)\sim F_U,\\
&\bs X = f_{\bs X}(U_{\bs X}),\\
&A = f_A(\bs X,U_A),\\
&Y = f_Y(A,\bs X,U_Y),
\end{align*}
where $f_{\bs X}, f_A, f_Y$ are left unrestricted and $U_{\bs X}, U_A, U_Y$ are mutually independent. Under this model, together with the identification assumptions above, we avoid committing to a particular parametric form for $f_{\bs X}, f_A,$ or $f_Y$. As a result, the ATE remains identified by the same g-formula functional, but estimation targets this functional directly rather than a parametric path coefficient.

TMLE leverages the orthogonal factorization of the observed-data likelihood,
\begin{align*}
\mathcal{L}(Y,A,\bs X)
= \mathbb{P}(Y\mid A,\bs X)\,\mathbb{P}(A\mid \bs X)\,\mathbb{P}(\bs X),
\end{align*}
and defines
\begin{align*}
Q(A,\bs X)=\mathbb{E}(Y\mid A,\bs X),
\qquad
g(A\mid \bs X)=\mathbb{P}(A\mid \bs X),
\end{align*}
where $Q$ is the outcome regression and $g$ is the propensity score, which is considered as a nuisance parameter and will be used in the targeting step. For concreteness, we assume $Y$ is continuous. In practice, one often bounds $Y$ to improve numerical stability; for an unbounded continuous outcome, one convenient option is min--max scaling \citep{van-der-laan-2011, frank-2023},
\begin{align*}
\tilde{Y}_i = \frac{Y_i-\min_j Y_j}{\max_j Y_j-\min_j Y_j}\in[0,1],
\end{align*}
followed by transforming estimates back to the original scale. The TMLE for ATE then involves the following three steps
\paragraph{Step 1: Initial estimation.}
Obtain initial estimates $Q_n^0(A,\bs X)$ for $Q(A,\bs X)$ and $g_n(A\mid \bs X)$ for $g(A\mid \bs X)$ using parametric regression or flexible machine learning with cross-validation, such as the Super Learner. The prediction from the fitted model yields $Q_n^0 (A = 1, \bs X_i)$ and $Q_n^0 (A = 0, \bs X_i), \forall i \in \{1,2, \cdots, n\}$. These predictions can be used to construct the g-computation, but TMLE will continue to the targeting step.

This step provides a first estimate of the outcome mechanism and the treatment assignment mechanism. In particular, $Q_n^0$ captures how the outcome is expected to vary across treatment levels and covariates, while $g_n$ describes how treatment is allocated given $\bs X$. Although these initial estimates may already be reasonable, they are not yet tailored to the target parameter, and the following targeting step for update is crucial.
\paragraph{Step 2: Targeting via the EIF}
We first define the clever covariate
\begin{align*}
H(g_n,A,\bs X)
=
\frac{A}{g_n(A = 1\mid \bs X)}-\frac{1-A}{g_n(A = 0\mid \bs X)}.
\end{align*}
The fluctuation parameter can then be modeled via the logistic submodel, which is appropriate when $Y$ is bounded in $[0,1]$,
\begin{align*}
\text{logit}\!\left(Q_n^{\epsilon}(A,\bs X)\right)
=
\text{logit}\!\left(Q_n^0(A,\bs X)\right)+\epsilon\,H(g_n,A,\bs X),
\end{align*}
The estimation of $\hat{\epsilon}$ can then be done by fitting a logistic regression of $Y$ on $H(g_n,A,\bs X)$ with offset $\text{logit}(Q_n^0(A,\bs X))$. Thus, the targeting step will update the initial estimates to be
\begin{align*}
Q_n^*(A,\bs X_i)=Q_n^{\hat{\epsilon}}(A,\bs X_i) = \text{expit}\left( \text{logit}(Q^0_n (A, \bs X_i)) + \hat{\epsilon} H(g_n, A, \bs X_i)\right).
\end{align*}

Intuitively, the targeting step makes a principled correction to the initial outcome estimate $Q_n^0$. The clever covariate determines the direction in which $Q_n^0$ should be updated so that the resulting estimator better respects the efficient influence function equation. In this sense, TMLE does not discard the initial fit, but rather refines it in a way that is directly aligned with efficient estimation of the ATE.

\paragraph{Step 3: Plug-in estimation and inference.}
After obtaining the updated estimates, the final TMLE follow the targeting step for ATE is the substitution estimator
\begin{align*}
\hat{\psi}_{\textup{ATE}}^{\textup{TMLE}}
=
\frac{1}{n}\sum_{i=1}^n
\left\{Q_n^*(1,\bs X_i)-Q_n^*(0,\bs X_i)\right\}.
\end{align*}
Intuitively, once the updated regression $Q_n^*$ has been targeted toward the parameter of interest, the ATE can be estimated by simply averaging the predicted counterfactual contrast over the sample. The efficient influence function then provides a convenient way to quantify the remaining sampling variability, which leads to standard errors, confidence intervals, and hypothesis tests \citep{wasserman-2006}.
In particular, the EIF for ATE used for the targeting step is
\begin{align*}
D^*(O;Q,g,\psi)
=
H(g,A,\bs X)\{Y-Q(A,\bs X)\}+Q(1,\bs X)-Q(0,\bs X)-\psi,
\end{align*}
and an empirical estimate is obtained by plugging in $(Q_n^*,g_n,\hat{\psi}_{\textup{ATE}}^{\textup{TMLE}})$
\begin{align*}
\hat{D}(O_i)
=
H(g_n,A_i,\bs X_i)\{Y_i-Q_n^*(A_i,\bs X_i)\}
+Q_n^*(1,\bs X_i)-Q_n^*(0,\bs X_i)
-\hat{\psi}_{\textup{ATE}}^{\textup{TMLE}}.
\end{align*}
Thus, a consistent variance estimator is the variance of the efficient influence function
\begin{align*}
\widehat{\mathrm{Var}}\!\left(\hat{\psi}_{\textup{ATE}}^{\textup{TMLE}}\right)
=
\frac{1}{n} \cdot \frac{1}{n}\sum_{i=1}^n \hat{D}(O_i)^2,
\end{align*}
yielding a $95\%$ confidence interval
$\hat{\psi}_{\textup{ATE}}^{\textup{TMLE}} \pm z_{0.975}\,\hat{\sigma}/\sqrt{n}$,
where $\hat{\sigma}^2=\frac{1}{n}\sum_{i=1}^n \hat{D}(O_i)^2$ \citep{gruber-2009}. 
TMLE is known for its double robustness: if either $Q_n^0(A,\bs X)$ or $g_n(A\mid \bs X)$ is consistent, then $\hat{\psi}_{\textup{ATE}}^{\textup{TMLE}}$ is consistent for $\psi_{\textup{ATE}}$ \citep{van-der-laan-2011}. Moreover, when both outcome and nuisance estimators are consistent, TMLE attains the semiparametric efficiency bound.
Compared with SEM and path analysis, TMLE avoids committing to a specific parametric form for the structural relations; instead, it conducts estimation and inference within an NPSEM by targeting the causal effects directly as functionals of the observed-data distribution. When the working models are correctly specified (e.g., a linear outcome regression with appropriate interactions), the initial estimate $Q_n^0$ can be obtained from the same regression system used in path analysis. In such cases, the TMLE targeting update is asymptotically negligible, and the TMLE and path-analysis estimators are asymptotically equivalent. Under misspecification, however, the path-analysis estimator generally remains biased, whereas the targeting step yields a doubly robust estimator that can reduce bias in the target functional.

TMLE-based CATE estimation can be carried out by applying the same procedure within strata (or by targeting conditional parameters directly). Practical implementations are available in \texttt{tmle3} \citep{Jeremy-2021} and \texttt{tmle} \citep{gruber-2012}, which can integrate with \texttt{SuperLearner}. See \citet{frank-2023, luquefernandez-2018} for accessible implementation tutorials.

\subsection{Causal Mediation Analysis}
\subsubsection{Estimands and identification assumptions}
Causal mediation analysis decomposes the total effect of a treatment $A$ on an outcome $Y$ into a direct effect (not operating through a mediator) and an indirect effect (operating through a mediator $M$). Let $A\in\{0,1\}$ denote the treatment, $\bs W\in\mathcal{W}\subseteq\mathbb{R}^p$ baseline covariates, $M\in\mathcal{M}$ a mediator, and $Y$ a continuous outcome. Let $M(a)$ denote the potential mediator value under $A=a$, and $Y(a,m)$ denote the potential outcome under $(A=a,M=m)$. The composition assumption implies $Y(a,M(a))=Y(a)$ \citep{ding-2024}, and the total effect (ATE) can be written as
\begin{align*}
\psi_{\textup{ATE}}=\mathbb{E}\!\left[Y(1)-Y(0)\right]
=
\mathbb{E}\!\left[Y(1,M(1))-Y(0,M(0))\right].
\end{align*}
The natural direct effect (NDE) and natural indirect effect (NIE) are defined as
\begin{align*}
\text{NDE} &= \mathbb{E}\!\left[Y(1,M(0))-Y(0,M(0))\right],\\
\text{NIE} &= \mathbb{E}\!\left[Y(1,M(1))-Y(1,M(0))\right],
\end{align*}
and satisfy $\psi_{\textup{ATE}}=\text{NDE}+\text{NIE}$ under composition.
To identify the NIE and NDE, we need several assumptions in addition to the standard consistency assumption: {\it(i) sequential ignorability: } $A \perp Y(a, m) | \bs W$ and $M \perp Y(a, m) | \bs W, A$ for all $a \in \mathcal{A}, m \in \mathcal{M}$; {(ii) no treatment-mediator confounding: }$A \perp M(a) | \bs W$ for all $a \in \mathcal{A}$; and {\it (iii) cross-world independence:} $Y(a, m) \perp M(a') | \bs W$ for all $a, a' \in \mathcal{A}, m \in \mathcal{M}$ \citep{ding-2024}. The positivity assumption here is extended to be: $\exists \epsilon_a > 0 \text{ such that }\bP(A = a \mid \bs W = \bs w) > \epsilon_a$ $,\forall a \in \mathcal{A}, \bs w \in \mathcal{W}$ and $\exists \epsilon_m > 0 \text{ such that } \bP(M = m \mid A = a, \bs W = \bs w) > \epsilon_m$ $\forall m \in \mathcal{M}, a \in \mathcal{A}, \bs w \in \mathcal{W}$. Under these assumptions, NDE and NIE can be identified as
\begin{align*}
\text{NDE}
&=
\mathbb{E}_{\bs W}
\left[
\mathbb{E}_{M\mid A=0,\bs W}
\left\{
\mathbb{E}(Y\mid A=1,M,\bs W)-\mathbb{E}(Y\mid A=0,M,\bs W)
\right\}
\right],\\
\text{NIE}
&=
\mathbb{E}_{\bs W}
\left[
\mathbb{E}_{M\mid A=1,\bs W}\!\left\{\mathbb{E}(Y\mid A=1,M,\bs W)\right\}
-
\mathbb{E}_{M\mid A=0,\bs W}\!\left\{\mathbb{E}(Y\mid A=1,M,\bs W)\right\}
\right].
\end{align*}
In the following subsections we highlight the analytic equivalence between path analysis and TMLE under correct parametric specification, and the robustness advantages of TMLE under misspecification.

\subsubsection{Estimating NDE and NIE via path analysis}
\setlength{\abovedisplayskip}{4pt}
\setlength{\belowdisplayskip}{4pt}

A common approach in SEM-based mediation analysis is to encode the identifying conditional expectations using a parametric path model \citep{de-stavola-2014}. Consider the linear structural equations
\begin{align*}
&U=(U_{\bs W},U_A,U_M,U_Y)\sim P_U,\\
&\bs W = U_{\bs W},\\
&A = g_A(\bs W,U_A),\\
&M = \alpha A + \bs\Gamma^\top \bs W + U_M,\\
&Y = \gamma A + \beta M + \bs\Theta^\top \bs W + U_Y,
\end{align*}
where exogenous terms are mutually independent, and $g_A$ is a known function. Under these parametric assumptions, the natural effects admit closed-form expressions \citep{gunzler-2013, ding-2024}.

\begin{figure}[h]
\centering
\begin{tikzpicture}
\node[draw, circle, minimum size=0.7cm, inner sep=2pt] (W) at (0,0) {$\bs W$};
\node[draw, circle, minimum size=0.7cm, inner sep=2pt] (A) at (2,0) {$A$};
\node[draw, circle, minimum size=0.7cm, inner sep=2pt] (M) at (4,0) {$M$};
\node[draw, circle, minimum size=0.7cm, inner sep=2pt] (Y) at (6,0) {$Y$};
\draw[->] (W) -- (A);
\draw[->] (A) -- (M);
\draw[->] (M) -- (Y);
\draw[->] (W) to [out=60,in=120] (M);
\draw[->] (A) to [out=-60,in=-120] (Y);
\draw[->] (W) to [out=-60,in=-120] (Y);
\end{tikzpicture}
\caption{Causal directed acyclic graph depicting baseline covariates $\bs W$, exposure $A$, mediator $M$, and outcome $Y$.}
\label{fig:med_dag}
\end{figure}

In particular, we may identify the NDE to be
\begin{align*}
    \text{NDE} &= \bE_{\bs W} \left[ \bE_{M|A = 0, \bs W} \left( \bE(Y |A = 1, M, \bs W) - \bE(Y |A = 0, M, \bs W) \right) \right]\\
    & =\sum_{\bs w \in \mathcal{W}} \int_{m \in \mathcal{M}} \left[ \mathbb{E}(Y | A = 1, M = m, \bs W = \bs w) - \mathbb{E}(Y | A = 0, M = m, \bs W = \bs w) \right] \\
     & \times f(M = m | A = 0, \bs W = \bs w) \mathrm{d} m \mathbb{P}( \bs W = \bs w)\\
    & = \sum_{\bs w \in \mathcal{W}} \int_{m \in \mathcal{M}} \left( \gamma + \beta m + \bs \Theta^\top \bs w - \beta m - \bs \Theta^\top \bs w \right) \times f(M = m | A = 0, \bs W = \bs w) \mathrm{d} m \mathbb{P}(\bs W = \bs w)\\
    & = \sum_{\bs w \in \mathcal{W}} \int_{m \in \mathcal{M}} \gamma \times f(M = m | A = 0, \bs W = \bs w) \mathrm{d} m \mathbb{P}(\bs W = \bs w) = \sum_{\bs w \in \mathcal{W}} \gamma \mathbb{P}(\bs W = \bs w) = \gamma
\end{align*}
The NIE can be identified similarly by
\begin{align*}
    \text{NIE} & = \bE_{\bs W} \left[ \bE_{M|A = 1, \bs W}(\bE(Y |A = 1, M, \bs W)) - \bE_{M|A = 0, \bs W}(\bE(Y |A = 1, M, \bs W))\right]\\
    & = \sum_{\bs{w} \in \mathcal{W}} \int_{m \in \mathcal{M}} \mathbb{E}(Y | A = 1, M = m, \bs{W} = \bs{w}) \times \{f(M = m | A = 1, \bs{W} = \bs{w}) \\
    & - f(M = m | A = 0, \bs{W} = \bs{w})\} \mathrm{d} m \mathbb{P}(\bs{W} = \bs{w})\\
    & = \sum_{\bs{w} \in \mathcal{W}} \int_{m \in \mathcal{M}} (\gamma + \beta m + \bs \Theta^\top \bs{w})\times \{f(M = m | A = 1, \bs{W} = \bs{w}) \\
    & - f(M = m | A = 0, \bs{W} = \bs{w})\} \mathrm{d} m \mathbb{P}(\bs{W} = \bs{w})\\
    & = \sum_{\bs{w} \in \mathcal{W}} \beta(\mathbb{E}(M | A = 1, \bs{W} = \bs{w}) - \mathbb{E}(M | A = 0, \bs{W} = \bs{w}))\mathbb{P}(\bs{W} = \bs{w})\\
    &= \sum_{\bs{w} \in \mathcal{W}} \beta \alpha \mathbb{P}(\bs{W} = \bs{w}) = \beta \alpha.
\end{align*}
Therefore, under correct specification, estimating NDE and NIE is equivalent to estimating the relevant path coefficients in the parametric SEM. Inference for the direct effect can proceed via Wald tests, while inference for the indirect effect commonly uses the Delta method or bootstrap \citep{rosseel-2012}.

It is worth mentioning that beyond the identification assumptions above, expressing mediational effects as SEM path coefficients typically requires additional parametric restrictions on the structural equations. Under model misspecification, the estimator based on the path analysis may lose its desired properties. To relax these strong assumptions and reduce vulnerability to misspecification, TMLE targets the mediational effects directly as causal functionals of the observed-data distribution, yielding plug-in estimators with robustness and improved stability under model misspecification.

\subsubsection{Estimating NDE and NIE via TMLE}
TMLE again relaxes the parametric assumptions of path analysis by encoding the data-generating process as an NPSEM and targeting the mediation estimands directly as functionals of the observed-data distribution. Following \citet{zheng-2012}, we outline TMLE for the NDE while the TMLE for the NIE is obtained analogously.

Consider the NPSEM \citep{pearl-2009}
\begin{align*}
&U=(U_{\bs W},U_A,U_M,U_Y)\sim P_U,\\
&\bs W = f_{\bs W}(U_{\bs W}),\\
&A = f_A(\bs W,U_A),\\
&M = f_M(\bs W,A,U_M),\\
&Y = f_Y(\bs W,A,M,U_Y),
\end{align*}
where $f_{\bs W}, f_A, f_M, f_Y$ are left unrestricted and the exogenous errors $U_{\bs W},U_A,U_M,U_Y$ are mutually independent. This model implies the observed-data likelihood factorization
\begin{align*}
\mathcal{L}(O)
=
\mathbb{P}(\bs W)\,\mathbb{P}(A\mid \bs W)\,\mathbb{P}(M\mid A,\bs W)\,\mathbb{P}(Y\mid A,M,\bs W).
\end{align*}
Define the outcome and nuisance components
\begin{align*}
\bar{Q}_Y(A,M,\bs W) &= \mathbb{E}(Y\mid A,M,\bs W),\\
Q_M(M\mid A,\bs W) &= \mathbb{P}(M\mid A,\bs W),\\
g(A\mid \bs W) &= \mathbb{P}(A\mid \bs W).
\end{align*}
TMLE targets $\bar{Q}_Y$ and an intermediate mediated mean that integrates $\bar{Q}_Y$ over the mediator distribution under $A=0$. For simplicity, we assume the relevant conditional means are bounded in $[0,1]$ (or are transformed to be bounded) and use logistic fluctuation submodels for the targeting steps. Since the approach is analogous to that for the ATE, we provide only a general outline of the TMLE procedure for mediation analysis and refer the reader to \citet{zheng-2012} for detailed derivations and proofs.

\paragraph{Step 1: Initial estimation.}
Estimate $\hat{\bar{Q}}_Y(A,M,\bs W)$, $\hat{Q}_M(M\mid A,\bs W)$, and $\hat{g}(A\mid \bs W)$ using parametric regression or flexible machine learning with cross-validation.

\paragraph{Step 2: Target the outcome regression.}
Estimate $\epsilon_1$ by minimizing empirical cross-entropy risk under the working submodel
\begin{align*}
\bar{Q}_Y^{\epsilon_1}(A,M,\bs W)
=
\text{expit}\!\left(
\text{logit}\big[\hat{\bar{Q}}_Y(A,M,\bs W)\big]
+
\epsilon_1\,C_Y(\hat{Q}_M,\hat{g};A,M,\bs W)
\right),
\end{align*}
where the clever covariate is
\begin{align*}
C_Y(\hat{Q}_M,\hat{g};A,M,\bs W)
=
\frac{\mathbf{1}(A=1)}{\hat{g}(1\mid \bs W)}
\cdot
\frac{\hat{Q}_M(M\mid 0,\bs W)}{\hat{Q}_M(M\mid 1,\bs W)}
-
\frac{\mathbf{1}(A=0)}{\hat{g}(0\mid \bs W)}.
\end{align*}
Denote $\hat{\bar{Q}}_Y^*(A,M,\bs W)=\bar{Q}_Y^{\hat{\epsilon}_1}(A,M,\bs W)$, the updated estimates.

\paragraph{Step 3: Target the mediated mean functional.}
Using $\hat{\bar{Q}}_Y^*$, construct an initial estimator of the mediated conditional mean under $A=0$,
\begin{align*}
\hat{\mathbb{E}}_{M}\!\left[\hat{\bar{Q}}_Y^*(1,M,\bs W)-\hat{\bar{Q}}_Y^*(0,M,\bs W)\mid A=0,\bs W\right],
\end{align*}
and apply a second targeting update (with clever covariate $C_M(\hat{g})=1/\hat{g}(0\mid \bs W)$) as in \citet{zheng-2012} to obtain a targeted estimate $\hat{\mathbb{E}}_M^*(\bs W)$.

\paragraph{Step 4: Plug-in estimation and inference.}
The TMLE for NDE is the substitution estimator
\begin{align*}
\hat{\psi}_{\textup{NDE}}^{\textup{TMLE}}
=
\frac{1}{n}\sum_{i=1}^n \hat{\mathbb{E}}_M^*(\bs W_i),
\end{align*}
and inference follows from the corresponding EIF-based variance estimator \citep{zheng-2012, van-der-laan-2011}. TMLE for NIE is obtained analogously by targeting the appropriate mediation functional.

When the mediator and outcome models are correctly specified as linear parametric regressions, the initial estimators in TMLE can be taken from the corresponding path-analysis fits. In this case, the targeting updates are asymptotically negligible and TMLE is asymptotically equivalent to the standard SEM and path-analysis estimators of NDE and NIE. Under misspecification, however, path-analysis estimators generally remain biased, whereas TMLE can reduce bias through targeting and retain double robustness when key nuisance components are consistently estimated.

For software implementation, \citet{hejazi-2022} provides the \texttt{R} package \texttt{medoutcon} for estimating natural effects using TMLE with Super Learner libraries. An illustration of code implementation corresponding to the real data example considered in the Application Section is also given in the Appendix.

\section{Simulation Studies}

To illustrate (i) the asymptotic equivalence between path-analysis estimators and TMLE under correct parametric specification and (ii) the comparative robustness of TMLE under model misspecification, we conducted Monte Carlo simulation studies for both ATE/CATE and causal mediation effects. Across all experiments, we evaluate point estimation accuracy and the validity of corresponding uncertainty quantification. TMLE is implemented with Super Learner libraries to allow flexible initial estimation, while the path-analysis approach is implemented via standard parametric regression or SEM fits.  

\subsection{Average Treatment Effects: Methods}

\subsubsection{Data-generating process}
We adopt a data-generating process (DGP) similar to \citet{luquefernandez-2018}, which corresponds to the structural causal graph in Figure~\ref{fig:causalgraph_ate}. Baseline covariates $\bs W=(W_1,W_2,W_3,W_4)$ jointly affect both treatment assignment $A$ and outcome $Y$, inducing confounding that must be adjusted for in estimating causal effects.

\begin{figure}[htbp]
\centering
\begin{tikzpicture}[scale = 0.65, node distance=1.5cm and 2.5cm, every node/.style={draw, circle},
    >=Stealth, every path/.style={-Stealth}]
    \node (w1) at (0, 0) {$W_1$};
    \node (w2) at (0, 6) {$W_2$};
    \node (w3) at (4, 5) {$W_3$};
    \node (w4) at (4, 0.8) {$W_4$};
    \node (a) at (7, 3) {$A$};
    \node (y) at (10, 3) {$Y$};
    \draw (w1) -- (w3);
    \draw (w1) -- (a);
    \draw (w1) -- (w4);
    \draw (w2) -- (w3);
    \draw (w2) -- (w4);
    \draw (w2) to[bend left=45] (y);
    \draw (w3) -- (a);
    \draw (w3) -- (y);
    \draw (w4) -- (a);
    \draw (w4) -- (y);
    \draw (a) -- (y) [red];
    \draw (w1) to[bend right=45] (y);
\end{tikzpicture}
\caption{Structural causal graph adapted from \citet{luquefernandez-2018} for the ATE/CATE simulations. Baseline covariates $W_1$--$W_4$ confound the effect of treatment $A$ on outcome $Y$ (red arrow).}
\label{fig:causalgraph_ate}
\end{figure}
Specifically, the data-generating process is defined as 
\begin{align*}
&W_1 \sim \text{Bernoulli}(0.5), 
\qquad
W_2 \sim \text{Bernoulli}(0.65),\\
&W_3 \sim \text{Uniform}(\{0,1,2,3,4\}),
\qquad
W_4 \sim \text{Uniform}(\{0,1,2,3,4,5\}),\\
&A \mid \bs W \sim \text{Bernoulli}\!\left(\text{plogis}\!\left(-2.5 + 0.05\,W_2 + 0.25\,W_3 + 0.6\,W_4 + 0.4\,W_2W_4\right)\right),\\
&Y \mid A,\bs W \sim \mathcal{N}\!\left(-1 + \psi\,A + 0.1\,W_1 + 0.35\,W_2 + 0.25\,W_3 + 0.2\,W_4 + 3.0\,W_2W_4,\; 1\right),
\end{align*}
where $\psi \in \{0.5, 1.5\}$ is the true ATE. To showcase the robustness of the TMLE, we consider the following model specification:
\begin{itemize}
\item \textbf{Correct:} The working models for estimation match the DGP (including the $W_2W_4$ interaction in the outcome model).
\item \textbf{NoInteraction:} The DGP remains unchanged, but the working outcome model omits the interaction term $W_2W_4$ when estimating the ATE/CATE.
\item \textbf{NonLinear:} The DGP is modified to include nonlinear covariate effects in the outcome model:
\begin{align*}
Y \mid A,\bs W \sim \mathcal{N}\!\left(-1 + \psi\,A + 0.1\,W_1 + 0.35\,W_2 + 0.25\,W_3^4 + 0.2\,W_4^4 + 3.0\,W_2W_4,\; 1\right).
\end{align*}
\item \textbf{NonNormal:} The conditional mean model is correctly specified, but the outcome noise is heavy-tailed:
$Y \mid A,\bs W$ follows a Student-$t$ distribution with $2$ degrees of freedom and the same conditional mean.
\end{itemize}
In the following comparison, we consider the following two estimators
\begin{enumerate}
\item \textbf{Parametric path-analysis estimator (working regression/SEM):} For ATE, we fit a linear regression for $Y$ given $(A,\bs W)$ using \texttt{lm} in \texttt{R} and interpret the coefficient of $A$ as the ATE under correct specification. Equivalently, one could fit an SEM with a logit link for $A$ and a linear model for $Y$; results are analogous under the same working mean model.
\item \textbf{TMLE:} We apply TMLE in \texttt{R} using the \texttt{tmle} package \citep{gruber-2012}, with Super Learner libraries used to estimate both the outcome regression $Q(A,\bs W)$ and propensity score $g(A\mid \bs W)$.
\end{enumerate}
The Super Learner library includes generalized linear models (\texttt{glm}, \texttt{glm.interaction}), generalized additive models (\texttt{gam}), random forests (\texttt{ranger}), and gradient boosting (\texttt{xgboost}). These learners allow flexible estimation when nonlinearities or interactions are present.
We conduct Monte Carlo simulations with $n_{\text{sim}}=1000$ replications for sample sizes
\[
n \in \{200, 500, 750, 1000, 1500, 2000, 3000\}.
\]
For each replication, we compute: (i) \textbf{Relative Bias:} $\frac{1}{n_{\text{sim}}}\sum_{s=1}^{n_{\text{sim}}}\frac{\hat{\psi}^{(s)}-\psi}{\psi}$; (ii) \textbf{Standardized RMSE:} $\sqrt{\frac{1}{n_{\text{sim}}}\sum_{s=1}^{n_{\text{sim}}}\left(\frac{\hat{\psi}^{(s)}-\psi}{\psi}\right)^2}$; (iii) \textbf{95\% CI coverage:} $\frac{1}{n_{\text{sim}}}\sum_{s=1}^{n_{\text{sim}}}\mathbf{1}\{\psi\in(\ell^{(s)},u^{(s)})\}$; (iv) \textbf{Power against $H_0:\psi=0$:} $\frac{1}{n_{\text{sim}}}\sum_{s=1}^{n_{\text{sim}}}\mathbf{1}\{0\notin(\ell^{(s)},u^{(s)})\}$, where $(\ell^{(s)},u^{(s)})$ denotes the estimator-specific 95\% confidence interval in replication $s$.
To study treatment effect heterogeneity and consider the CATE, we modify the outcome DGP by adding an interaction between $A$ and $W_1$:
\begin{align*}
Y \mid A,\bs W \sim \mathcal{N}\!\left(
-1 + \psi\,A + 0.5\,A W_1 + 0.1\,W_1 + 0.35\,W_2 + 0.25\,W_3 + 0.2\,W_4 + 3.0\,W_2W_4,\; 1\right),
\end{align*}
and define the target estimand as the CATE among individuals with $W_1=1$. Both the parametric regression and TMLE are refit accordingly, and performance is evaluated using the same metrics.

\subsection{Mediation Analysis: Methods}
\subsubsection{Data-generating process}
For the mediation analysis, we generate i.i.d.\ data $(W,A,M,Y)$ from the following data-generating process
\begin{align*}
&W \sim \mathcal{N}(0,1),\\
&A\mid W \sim \text{Bernoulli}\!\left(\text{plogis}(0.5\,W)\right),\\
&M\mid A,W \sim \mathcal{N}\!\left(A + 0.5\,W,\;1\right),\\
&Y\mid A,W,M \sim \mathcal{N}\!\left(2A + M + 0.8\,W,\;1\right).
\end{align*}
Under this DGP, the true natural indirect effect and natural direct effect are $\psi_{\text{NIE}}=1$ and $\psi_{\text{NDE}}=2$, respectively.
In the following comparison, we again consider
\begin{enumerate}
\item \textbf{Parametric SEM/path analysis:} Implemented in \texttt{lavaan}. Uncertainty is quantified via bootstrap with 1000 resamples, consistent with common practice in SEM mediation analysis.
\item \textbf{TMLE for mediation:} Implemented using \texttt{medoutcon} \citep{hejazi-2022}. Super Learner libraries include generalized linear models and random forests to estimate the nuisance components required in the targeting steps.
\end{enumerate}
Due to computational costs, we run $n_{\text{sim}}=200$ replications for sample sizes
\[
n \in \{500, 800, 1000, 1500, 2000, 2500, 3000\}.
\]
Two model misspecifications are considered
\begin{itemize}
\item \textbf{MisspecYW:} Nonlinear $Y$--$W$ relationship:
\begin{align*}
Y \mid A,W,M \sim \mathcal{N}\!\left(2A + M + 0.8\,W^4,\;1\right).
\end{align*}
\item \textbf{MisspecMWYW:} Nonlinearities in both mediator and outcome models:
\begin{align*}
&M \mid A,W \sim \mathcal{N}\!\left(A + 0.5\,W^2,\;1\right),\\
&Y \mid A,W,M \sim \mathcal{N}\!\left(2A + M + 0.8\,W^4,\;1\right).
\end{align*}
\end{itemize}
Both estimators are refit under these scenarios, and performance is evaluated using the same metrics as in the ATE/CATE simulations, including relative bias, standardized RMSE, 95\% CI coverage, and power against $\psi=0$.

\subsection{Results for Average Treatment Effects}

\begin{figure}[ht]
\centering
\includegraphics[width=\linewidth]{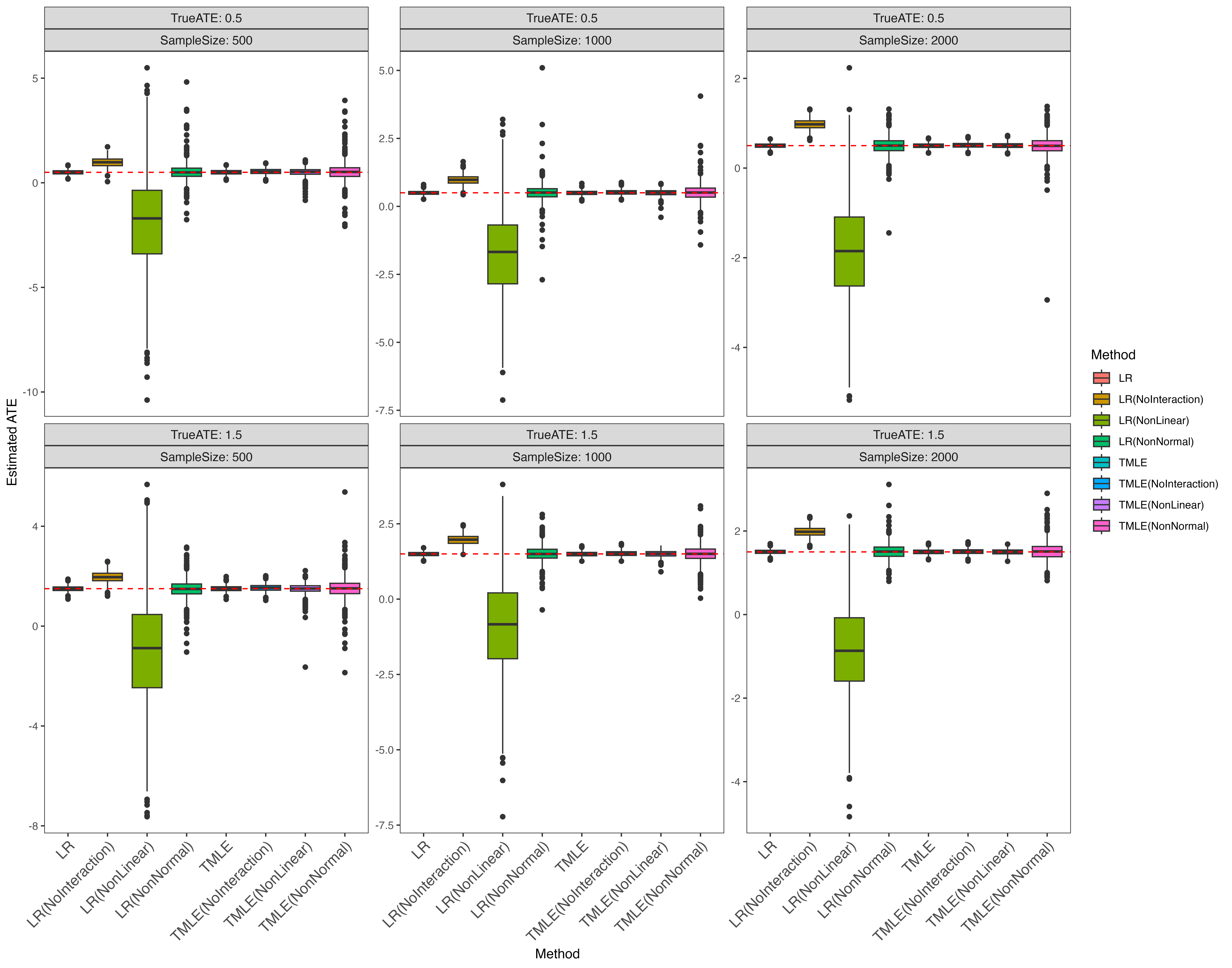}
\caption{Distributional boxplots for ATE across 1000 Monte Carlo replications.}
\label{fig:ate_box}
\end{figure}

\begin{figure}[htbp]
\centering
\includegraphics[width=\linewidth]{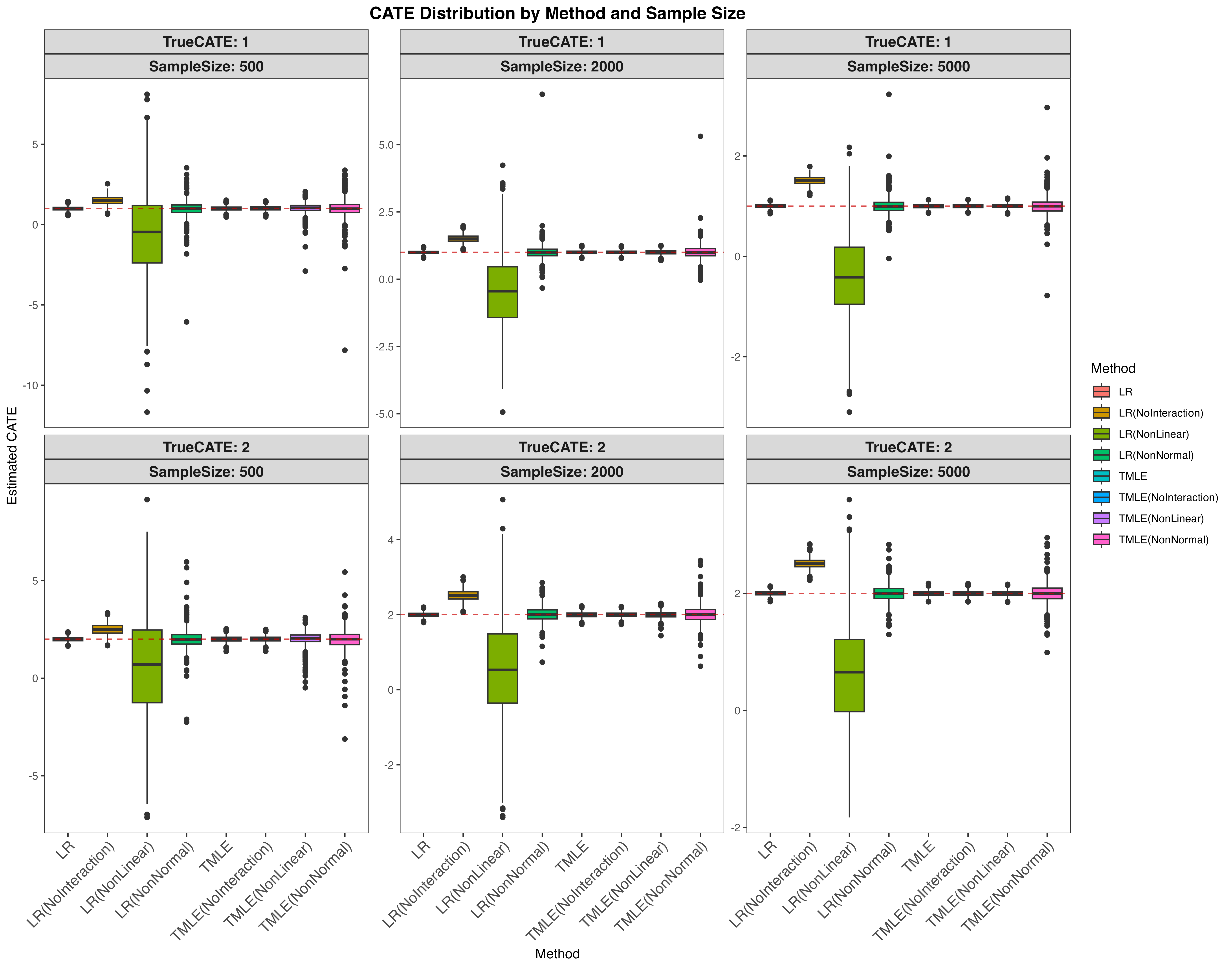}
\caption{Distributional boxplots for CATE (subgroup $W_1=1$) across 1000 Monte Carlo replications.}
\label{fig:cate_box}
\end{figure}

Figures~\ref{fig:ate_box} and \ref{fig:cate_box} summarize the distribution of Monte Carlo estimates for ATE and CATE across replications. Under correct specification, both the parametric regression from the path analysis working model and TMLE yield estimates centered near the true value with comparable spread, consistent with the equivalence of TMLE and parametric SEM when working models are correctly specified.

Under misspecification, the two approaches diverge significantly. In the \textbf{NoInteraction} and \textbf{NonLinear} settings, the regression-based estimator exhibits persistent bias even at large sample sizes (e.g., $n=3000$), reflecting convergence to a pseudo-true parameter under an incorrect working model. In contrast, TMLE remains approximately unbiased and centers around the true causal effect at substantially smaller sample sizes, consistent with its robustness proporties and the flexibility introduced by Super Learner.

Figures~\ref{fig:ate_metrics} and \ref{fig:cate_metrics} report performance across metrics as a function of sample size. When the model is correctly specified, both approaches show near-zero relative bias and standardized RMSE, and power approaches one as $n$ increases. Under misspecification, including omitted interactions and non-linear relations, path analysis exhibits declining CI coverage and reduced power due to biased point estimation, whereas TMLE maintains coverage close to the nominal 0.95 and achieves high power at moderate sample sizes. This is expected as the path analysis is sensitive to the correct specification while TMLE gains the robustness via the NPSEM. In the \textbf{NonNormal} setting, both methods remain broadly stable, with TMLE showing slightly improved finite-sample behavior in some metrics, though differences are modest.

\begin{figure}[htbp]
\centering
\includegraphics[width=\linewidth]{combined_simulation_plots.png}
\caption{Performance of the regression-based (parametric SEM working model) estimator and TMLE for ATE across metrics and sample sizes.}
\label{fig:ate_metrics}
\end{figure}

\begin{figure}[htbp]
\centering
\includegraphics[width=\linewidth]{combined_plot_CATE.png}
\caption{Performance of the regression-based estimator and TMLE for CATE (subgroup $W_1=1$) across metrics and sample sizes.}
\label{fig:cate_metrics}
\end{figure}
Overall, these results highlight the central message of this tutorial: when the parametric assumptions underlying SEM hold, path-analysis estimators and TMLE are empirically equivalent and typically yield similar results; however, when those assumptions fail, TMLE, based on NPSEM and combined with Super Learner, tends to provide more reliable point estimates and uncertainty quantification.

\subsection{Results for Mediation Analysis}

\begin{figure}[htbp]
\centering
\includegraphics[width=1.05\linewidth]{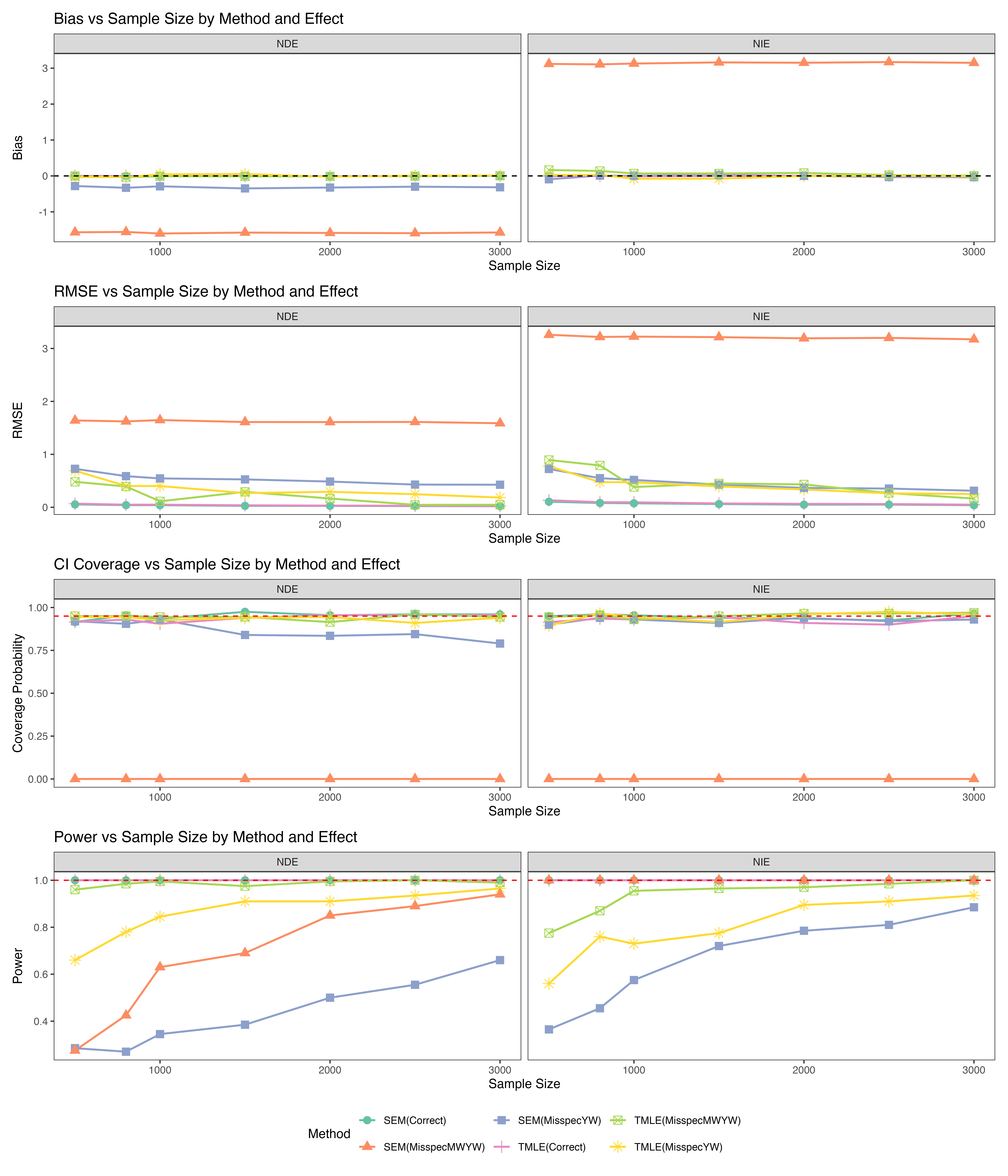}
\caption{Performance of SEM/path analysis and TMLE for NDE, NIE, and total effect across metrics and sample sizes under correct specification and misspecification scenarios.}
\label{fig:med_metrics}
\end{figure}

Figure~\ref{fig:med_metrics} summarizes performance for path analysis and TMLE across sample sizes and model scenarios. Under correct linear specification, both approaches yield unbiased estimates for NDE and NIE with decreasing RMSE and near-nominal confidence interval coverage, consistent with their theoretical equivalence when the parametric SEM is correctly specified.

Under misspecification, however, SEM becomes sensitive to incorrect functional forms. In \textbf{MisspecYW}, SEM exhibits noticeable bias in the NDE and degraded confidence interval coverage, with the NIE also impacted through the induced joint likelihood fit. In contrast, TMLE remains approximately unbiased and maintains coverage close to 0.95, even at moderate sample sizes (e.g., $n\approx 500$), reflecting the benefit of targeting and the use of flexible nuisance estimation.

The contrast becomes more pronounced in \textbf{MisspecMWYW}, where both mediator and outcome models deviate from linearity. SEM shows substantial bias for both NDE and NIE and confidence intervals that fail to achieve nominal coverage, whereas TMLE remains stable across both estimands, with improved RMSE and consistently valid confidence sets. TMLE also exhibits higher power under misspecification, indicating greater ability to detect nonzero mediation effects when the data-generating mechanism is complex or uncertain.

\section{Applications}

To demonstrate the practical utility of targeted learning for estimating ATE, CATE, and mediation effects in an applied setting, we analyze a real-world sociological question: the causal impact of multidimensional poverty on rural children’s access to high school education in China. We further examine gender heterogeneity and investigate whether children’s educational expenditure mediates the poverty--education relationship.

\subsection{Data and Study Design}

Our analysis draws on six waves (2012--2022) of the China Family Panel Study (CFPS), a nationally representative longitudinal survey. The analytic sample includes $2{,}881$ rural children aged 6--15 in the 2012 wave and newly sampled rural children aged 8--15 in the 2014 wave, who were subsequently followed to assess high school enrollment status at age 15 or 16 in later waves. The binary outcome is high school enrollment \citep{lei-2021}.

Following \citet{szaflarski-2019}, multidimensional poverty is constructed from nine indicators spanning three dimensions: health (nutrition, child mortality), education (parental years of schooling, children’s elementary school attendance), and living standards (cooking fuel, sanitation, drinking water, electricity, housing assets). Dimensions and indicators are equally weighted, and a child is classified as multidimensionally poor if the weighted deprivation score is at least 0.33.

To support conditional exchangeability (unconfoundedness) for the effect of poverty on high school enrollment, we adjust for a rich set of baseline covariates measured in each child’s first observed wave, including: individual characteristics (age, gender, siblings), household and parental factors (household registration, living arrangements, parental education and occupation, political identity, $\log$ per-capita income, Dibao receipt), school quality, and province fixed effects. In the mediation analysis, we consider children’s educational expenditure as a mediator, operationalized in two ways: (i) the logarithm of absolute educational expenditure; and (ii) the proportion of household income spent on education.

\subsection{Estimation of Average and Conditional Average Treatment Effects}
We treat multidimensional poverty status as a binary exposure $A\in\{0,1\}$ and high school enrollment as the binary outcome $Y$. The primary estimand is the ATE,
\[
\psi_{\textup{ATE}}=\mathbb{E}[Y(1)-Y(0)],
\]
interpreted as the average causal effect of being multidimensionally poor on high school enrollment. To assess gender heterogeneity, we estimate CATEs stratified by gender, corresponding to subgroup-specific average effects.
We compare a conventional parametric approach and TMLE:
\begin{enumerate}
\item \textbf{Parametric logistic regression:} We fit a logistic regression using \texttt{glm} in \texttt{R} with $Y$ regressed on $A$ and baseline covariates. The resulting coefficient on $A$ is transformed to a marginal effect scale and reported as an ATE estimate for comparability.
\item \textbf{TMLE:} We implement TMLE using \texttt{tmle} \citep{gruber-2012}. Initial estimates for the outcome regression and propensity score are obtained using the default Super Learner library in \texttt{tmle}, which includes regularized generalized linear models and Bayesian additive regression trees, among others.
\end{enumerate}
\subsubsection{Results: overall ATE}
Table~\ref{tab:ate_overall} reports the estimated average impact of multidimensional poverty on high school enrollment. TMLE yields an ATE of $-0.106$ (95\% CI: $-0.164$ to $-0.048$), while the parametric logistic regression yields a larger negative estimate of $-0.140$ (95\% CI: $-0.230$ to $-0.051$). Both approaches indicate that multidimensional poverty reduces the probability of high school enrollment, and both effects are statistically significant at $\alpha=0.05$.
\begin{table}[htbp]
\centering
\small
\renewcommand{\arraystretch}{0.9}
\caption{Average impact of multidimensional poverty on children’s high school enrollment.}
\label{tab:ate_overall}
\begin{tabular}{lcc}
\toprule
& TMLE & LOGIT \\
\midrule
ATE & $-0.106$ & $-0.140$ \\
S.E. & $0.030$ & $0.046$ \\
95\% CI & $(-0.164,\,-0.048)$ & $(-0.230,\,-0.051)$ \\
\bottomrule
\end{tabular}
\end{table}

\subsubsection{Results: gender-stratified CATE}
Table~\ref{tab:ate_gender} reports gender-stratified effects. For males, TMLE estimates an ATE of $-0.169$ (95\% CI: $-0.249$ to $-0.089$), whereas logistic regression estimates $-0.147$ (95\% CI: $-0.275$ to $-0.018$). For females, TMLE estimates a smaller negative effect of $-0.084$ (95\% CI: $-0.153$ to $-0.016$), while logistic regression estimates $-0.134$ (95\% CI: $-0.256$ to $-0.011$). Both methods suggest larger adverse effects of poverty on high school enrollment for males than for females, with TMLE indicating a more pronounced gender contrast. All subgroup effects are statistically significant at $\alpha=0.05$.

\begin{table}[ht]
\centering
\small
\renewcommand{\arraystretch}{0.9}
\caption{Average impact of multidimensional poverty on high school enrollment by gender.}
\label{tab:ate_gender}
\begin{tabular}{lcccc}
\toprule
& \multicolumn{2}{c}{TMLE} & \multicolumn{2}{c}{LOGIT} \\
\cmidrule(lr){2-3}\cmidrule(lr){4-5}
& Male & Female & Male & Female \\
\midrule
ATE & $-0.169$ & $-0.084$ & $-0.147$ & $-0.134$ \\
S.E. & $0.041$ & $0.035$ & $0.066$ & $0.063$ \\
95\% CI & $(-0.249,\,-0.089)$ & $(-0.153,\,-0.016)$ & $(-0.275,\,-0.018)$ & $(-0.256,\,-0.011)$ \\
\bottomrule
\end{tabular}
\end{table}

These heterogeneous effects are broadly consistent with prior literature suggesting that boys’ educational outcomes may be more sensitive to family disadvantage during key educational transitions \citep{autor-2019, owens-2016, lei-2020}. One potential interpretation is that households often place higher educational expectations on boys, and when poverty constrains economic and parental resources, boys may experience larger declines in academic support, motivation, or persistence, increasing the risk of not progressing to high school relative to girls.

\subsection{Mediation Analysis}
Beyond average effects, sociological explanations often emphasize \emph{mechanisms}. We therefore examine whether educational expenditure mediates the effect of multidimensional poverty on high school enrollment. Using educational expenditure $M$ as a mediator, estimate the direct and indirect effects and compare: 
\begin{enumerate}
\item \textbf{SEM/path analysis:} Implemented using \texttt{lavaan} \citep{schumacker-2015}. Uncertainty is quantified using bootstrap-based confidence intervals.
\item \textbf{TMLE for mediation:} Implemented using \texttt{medoutcon} \citep{hejazi-2022}. To accommodate potential nonlinearity and sparsity, we specify a Super Learner library including random forests, LASSO, gradient boosting, and generalized additive models.
\end{enumerate}
We estimate mediation effects under two operationalizations of educational expenditure: log absolute spending and the expenditure-to-income ratio.
\subsubsection{Results}
Table~\ref{tab:mediation} reports the decomposed mediation effects. Under SEM/path analysis, both the direct and indirect effects are statistically significant across both mediator definitions, with relatively small estimated indirect effects. In contrast, TMLE yields statistically significant indirect effects but direct effects that are no longer statistically significant at $\alpha=0.05$. Specifically, when the mediator is log expenditure, SEM estimates a direct effect of $-0.097$ (95\% CI: $-0.187$ to $-0.007$) and an indirect effect of $-0.023$ (95\% CI: $-0.037$ to $-0.009$), whereas TMLE estimates a direct effect of $-0.150$ (95\% CI: $-0.302$ to $0.002$) and an indirect effect of $-0.110$ (95\% CI: $-0.216$ to $-0.003$). A similar pattern holds when the mediator is the expenditure-to-income ratio.

\begin{table}[h]
\centering
\footnotesize
\renewcommand{\arraystretch}{0.9}
\caption{Mediating effect decomposition of multidimensional poverty on high school enrollment.}
\label{tab:mediation}
\begin{tabular}{lcccc}
\toprule
& \multicolumn{2}{c}{Indirect Effect (NIE)} & \multicolumn{2}{c}{Direct Effect (NDE)} \\
\cmidrule(lr){2-3} \cmidrule(lr){4-5}
& Standardized Effect & 95\% CI & Standardized Effect & 95\% CI \\
\midrule
Model 7 (SEM; log spending) & $-0.023$ & $(-0.037,\,-0.009)$ & $-0.097$ & $(-0.187,\,-0.007)$ \\
Model 8 (SEM; ratio)        & $-0.020$ & $(-0.034,\,-0.006)$ & $-0.100$ & $(-0.192,\,-0.008)$ \\
Model 9 (TMLE; log spending)& $-0.110$ & $(-0.216,\,-0.003)$ & $-0.150$ & $(-0.302,\,0.002)$ \\
Model 10 (TMLE; ratio)      & $-0.105$ & $(-0.198,\,-0.013)$ & $-0.142$ & $(-0.296,\,0.011)$ \\
\bottomrule
\end{tabular}
\end{table}

The divergence between SEM and TMLE is substantively meaningful: although both approaches indicate that poverty reduces educational expenditure and that expenditure is associated with high school enrollment, TMLE suggests that the \emph{indirect} pathway through educational expenditure may account for a larger share of the total effect, while the remaining direct effect becomes less distinguishable from zero once flexible nuisance estimation and targeting are used. One plausible explanation is that the parametric SEM working models may be misspecified, which can induce biased decomposition and overly optimistic confidence intervals. This pattern is consistent with our simulation results, where path-analysis estimators were sensitive to functional form misspecification, whereas TMLE remained stable due to its targeting step and Super Learner-based initial estimation.

\section{Discussion}

This tutorial and study clarify both the conceptual connections and practical differences between Path Analysis within structural equation modeling (SEM) and Targeted Maximum Likelihood Estimation (TMLE) for causal inference. We framed path coefficients in parametric SEM and causal functionals identified under nonparametric structural equation models (NPSEM) as two representations of the same underlying objective: estimating well-defined causal parameters such as the average treatment effect (ATE), conditional average treatment effect (CATE), and natural direct and indirect effects. While both approaches rely on explicit causal assumptions for identification, they differ fundamentally in their reliance on parametric functional forms and their robustness to model misspecification. Through theoretical exposition, simulation studies, and a real-world application, we illustrated when these approaches coincide and when they diverge.

\subsection{Key findings}

Across simulations and empirical analyses, three main findings emerged. First, when the outcome, treatment, mediator, and covariate relationships were correctly specified within linear parametric models, SEM-based path analysis and TMLE produced nearly identical point estimates and valid uncertainty quantification for ATE, CATE, and mediation effects. This empirical equivalence reflects the theoretical result that TMLE reduces to the corresponding parametric estimator when the working models are correctly specified and the targeting step becomes asymptotically negligible.

Second, under a range of plausible misspecifications, including omitted interaction terms, nonlinear covariate effects, and non-normal outcome distributions, TMLE consistently retained approximate unbiasedness and near-nominal confidence interval coverage. This robustness arises from TMLE’s double robustness property and the explicit targeting of the estimand through the efficient influence function, coupled with flexible nuisance estimation via Super Learner. In contrast, SEM-based linear estimators converged to pseudo-true parameters under misspecification, leading to biased point estimates, diminished statistical power, and severely degraded confidence interval coverage.

Third, in the mediation setting, TMLE maintained valid inference even when the outcome or mediator mechanisms departed from linearity, whereas parametric SEM was highly sensitive to such deviations. The empirical application further highlighted that TMLE attributed a larger share of the poverty effect to indirect pathways through educational expenditure, while SEM suggested persistent direct effects—an inconsistency likely driven by parametric model constraints. Together, these results position TMLE as a practically robust complement to SEM when functional form uncertainty is non-negligible.

\subsection{Scope and limitations}

Our analysis deliberately focused on \emph{path analysis}, that is, SEM with only observed variables in the structural component. We did not address latent-variable SEM involving measurement models, factor loadings, or multiple-indicator constructs. Consequently, several important issues remain beyond the scope of this paper: (i) bias induced by measurement error in observed proxies for treatments, mediators, or outcomes; (ii) interactions between measurement model misspecification and structural path estimation; and (iii) emerging approaches that integrate targeted learning with latent-variable frameworks. Extending targeted learning methods to accommodate latent constructs and jointly model measurement and structural components is an important direction for future research, particularly in psychological and sociological applications where key variables are rarely observed without error.

\subsection{When to prefer which approach}

SEM and TMLE serve complementary roles rather than competing paradigms. SEM remains particularly attractive when latent constructs are central to the research question, when theory supports low-dimensional linear structures, and when explicit parametric constraints (e.g., equality constraints across groups or time points) are substantively meaningful. Its close integration with measurement theory and well-established diagnostic tools makes it indispensable in many applied domains.

TMLE, by contrast, is preferable when the functional form of relationships is uncertain, when treatment effect heterogeneity is anticipated, when mediators and outcomes may interact nonlinearly with covariates, or when valid inference under plausible misspecification is prioritized. In such settings, TMLE’s reliance on flexible nuisance estimation and its targeting of the causal estimand directly provide a safeguard against common modeling errors that can undermine parametric SEM analyses.

\subsection{Extensions and open directions}

Several promising extensions warrant further investigation. First, \emph{longitudinal} causal settings with time-varying treatments, mediators, and confounders naturally align with sequential TMLE and marginal structural models, whereas SEM analogs often rely on cross-lagged or latent growth structures; systematic comparisons in these contexts would be valuable. Second, \emph{missing data} mechanisms can be incorporated within targeted learning via inverse-probability weighting and augmented estimators, while SEM approaches typically rely on full-information maximum likelihood under strong assumptions about missingness. Third, \emph{interference and spillover effects}, common in social networks and educational settings, violate the standard SUTVA assumption; recent developments in network TMLE provide potential solutions not easily accommodated in traditional SEM. Fourth, \emph{transportability and generalizability} across populations may benefit from combining TMLE with reweighting or covariate shift adjustments, while SEM approaches could impose structured equality constraints across groups. Finally, advances in design-based targeted learning—including collaborative TMLE, debiased machine learning, and targeted regularization—offer opportunities to further stabilize finite-sample performance.

\subsection{Conclusion}

Path Analysis within SEM and TMLE represent two analytically connected yet methodologically distinct approaches to causal inference. When linear models grounded in substantive theory are credible, path analysis remains efficient, interpretable, and computationally convenient. When model misspecification is a realistic concern, as is often the case in complex social and behavioral data, TMLE offers a robust alternative that directly targets causal parameters and preserves valid inference through flexible estimation and updating based on the efficient influence function. By placing causal identification at the forefront and choosing either a well specified parametric SEM or a targeted learning framework with adaptive learners, applied researchers can substantially enhance the credibility of causal conclusions in the social and behavioral sciences.

\newpage

\printbibliography

\newpage

\appendix
\section{R Code for Statistical Analysis}

The following R code performs the average treatment effect estimation, conditional average treatment effects, and mediation analyses using TMLE and path analysis, which corresponds to the case study considered in the Applications Section.

\lstset{style=Rstyle}
\begin{lstlisting}
## Package needed for TMLE
library(tmle)
library(SuperLearner)
library(medoutcon)
library(sl3)
path <- "Path"
data <- haven::read_dta(path)
## Assuming that data is well-structured and ready to be analyzed
## Specifying the set of confounders
W <- subset(data, select = c(
  gender, onlychild, rhukou, age, withparents, logdisppc,
  party, isei_max, dibao, edu_max,
  keysch, pvcode
))
## Specify the Super Learners libraries to obtain the initial estimates
learners <- c("SL.xgboost", "SL.gam","SL.ranger", "SL.earth")
## Call the TMLE
tmle_ATE <- tmle(
  Y = data$gaozhong,         ## Y: outcome
  A = data$mpi,              ## A: binary exposure/treatment
  W = W,                     ## W: confounders
  Q.SL.library = learners,   ## Super Learner for Q(A,W)
  g.SL.library = learners,   ## Super Learner for g(W)
  family = "binomial"
)
## Output the estimates
tmle_ATE$estimates$ATE$psi ## This is the point estimates
tmle_ATE$estimates$ATE$CI  ## This is the corresponding 95%-CI

## The CATE for girls (gender == 0)
## Estimate the treatment effect within a subgroup.
W1 <- W %>% filter(data$gender == 0); W1$gender <- NULL
data1 <- data %>% filter(gender == 0)
tmle_cate1 <- tmle(
  Y = data1$gaozhong,
  A = data1$mpi,
  W = W1,
  family = "binomial"
)
summary(tmle_cate1)

## Using logistic regression 
logit_model <- glm(
  gaozhong ~ mpi + gender + onlychild + rhukou + age + withparents + edu_max +
    logdisppc + party + isei_max + dibao + keysch + pvcode,
  data = data,
  family = binomial(link = "logit")
)
## The causal effects is the model coefficients
summary(margins(logit_model1))

## The Code for mediation analysis using tmle
## Set up the learners
fglm_lrnr <- Lrnr_glm_fast$new()
earth_lrnr <- Lrnr_earth$new()
rf_lrnr <- Lrnr_ranger$new(num.trees = 200)
xg_lrnr <- Lrnr_xgboost$new()
lasso_lrnr <- Lrnr_glmnet$new(alpha = 1, nfolds = 5)
lrnr_lib <- Stack$new(earth_lrnr,xg_lrnr, lasso_lrnr)
sl_lrnr <- Lrnr_sl$new(learners = lrnr_lib)
## Baseline covariates used in mediation adjustment
covars <- c("gender", "onlychild", "rhukou", "age", "withparents",
            "logdisppc", "party", "isei_max", "dibao", "edu_max", "keysch")

## Using TMLE to estimate the NDE
tmle_med1_direct <- medoutcon::medoutcon(
  W = data[covars],            ## baseline confounders
  A = data$mpi,                ## treatment/exposure
  Z = NULL,                    ## intermediate confounders 
  M = data$logeduexpend_resid,  ## mediator 
  Y = data$gaozhong_resid,      ## outcome  
  g_learners = sl_lrnr, h_learners = sl_lrnr, b_learners = sl_lrnr,
  effect = "direct",
  estimator = "tmle"
)

## Estimate the NIE
tmle_med1_indirect <- medoutcon::medoutcon(
  W = data[covars],
  A = data$mpi,
  Z = NULL,
  M = data$logeduexpend_resid,
  Y = data$gaozhong_resid,
  g_learners = sl_lrnr, h_learners = sl_lrnr, b_learners = sl_lrnr,
  effect = "indirect",
  estimator = "tmle"
)
## Output the TMLE estimates
tmle_med1_indirect$theta ## This is the point estimates
tmle_med1_indirect$var ## This is the variance estimtes

## Using the lavaan for SEM and path analysis 
library(lavaan)
lm1 <- lm(gaozhong ~ factor(pvcode), data = data)
lm2 <- lm(logeduexpend ~ factor(pvcode), data = data)
data$gaozhong_resid   <- resid(lm1)
data$logeduexpend_resid <- resid(lm2)
SEMmodel1 <- '
  logeduexpend_resid ~ a*mpi + gender + onlychild + rhukou + age +
                       withparents + logdisppc + party + isei_max +
                       dibao + edu_max + keysch

  gaozhong_resid ~ b*logeduexpend_resid + c_prime*mpi + gender + onlychild +
                   rhukou + age + withparents + logdisppc + party +
                   isei_max + dibao + edu_max + keysch

  ind := a*b
  direct := c_prime
  total := c_prime + ind
'
SEMmed1 <- sem(
  model     = model1,
  data      = data,
  se        = "bootstrap",
  bootstrap = 1000
)
## Output the fitted lavaan model
summary(SEMmed1,
        standardized = TRUE,
        rsquare      = TRUE,
        fit.measures = TRUE)
\end{lstlisting}

\end{document}